\documentclass[twocolumn,amsmath,amssymb,aps,prb]{revtex4-2}
\usepackage{graphicx}
\usepackage{hyperref}
\usepackage{xcolor}
\usepackage{soul}
\begin{document}

\preprint{APS/123-QED}

\title{Valley caloritronics in a photodriven heterojunction of Dirac materials}

\author{Priyadarshini Kapri}
\email{pkapri@iitk.ac.in}
\author{Bashab Dey}
\author{Tarun Kanti Ghosh}

\affiliation{Department of Physics, Indian Institute of Technology-Kanpur, 
Kanpur-208 016, India }

\date{\today}

\begin{abstract}
We consider a lateral hetero-junction where the left and right leads are made of 
monolayer graphene and the middle region is made of a gapped tilted Dirac material 
(borophene or quinoid graphene) illuminated with off-resonant circularly polarized radiation. 
The tilt parameter $v_t$ makes the band gap indirect and smaller in magnitude as compared 
to Dirac materials without tilt. Exposure to radiation makes the band gaps 
of the central region valley-dependent which show their signatures as valley polarized 
charge and thermal currents, thereby causing a valley Seebeck effect. We study the variation 
of the valley polarized electrical conductance, thermal conductance, thermopower and figure of merit 
of this junction with chemical potential $\mu$ and a tunable gap parameter 
$\eta$. For non-zero $\eta$, all the valley polarized quantities are peaked 
at certain values of chemical potential and then vanish asymptotically. 
Increase in gap parameter enhances the valley thermopower and valley figure of merit, 
whereas the valley conductances (electrical and thermal) show non monotonic behavior with $\eta$. 
We also compare the valley polarized quantities with their 
corresponding charge counterparts (effective contribution from both the valleys). 
The charge thermopower and the charge figure of merit behave non monotonically 
with $\eta$ and the charge conductances (electrical and thermal) depict a decreasing 
trend with $\eta$. Furthermore, the tilt parameter reduces the effective transmission of 
carriers through the junction, thereby diminishing all the charge and valley polarized quantities.  
As the gaps in the dispersion  can be adjusted by varying the intensity of light 
as well as the Semenoff mass, the tunability of 
this junction with regard to its thermoelectric properties may be experimentally realizable. 

\end{abstract}

\maketitle

\section{Introduction}
Thermoelectric materials have attracted immense interest in energy efficient
device applications
\cite{Hicks1,DHicks1,Venkatasubramanian,Arita,Hamada,Zide,Wei,Zuev,Kato,
Buscema,Konabe,Hicks2,Dresselhaus}. 
The efficiency of  power generation in such devices depends on the interplay
between their electronic and thermal performances and it is characterized by a
dimensionless quantity called figure of merit, $ZT = S^2\sigma T/\kappa$, where
$\sigma$, $\kappa$ and $S$ denote electrical conductivity, thermal
conductivity and Seebeck
coefficient (thermopower), respectively, with $T$ being the absolute
temperature.
In bulk materials, the factors in the expression of $ZT$ are mutually
coupled in such a way that it is difficult to control them
independently,
and hence improve $ZT$. The techniques used to improve the
figure of merit rely on enhancing the power factor
($\sigma S^2$) and lowering the thermal conductivity.
One of the important proposals is
nanostructuring of materials which enhances thermoelectric efficiency
due to the sharply peaked density of states (DOS) of the carriers in low-dimensional
materials \cite{Hicks1,Bilc}. Another useful method is engineering the band
structure \cite{Heremans,Liu1} in conjunction with nanostructuring to lower
the thermal conductivity. Further, the use of semi-metals with large
electron-hole asymmetry
can enhance the thermoelectric coefficients \cite{Markov}.

Advancements in fabrication technologies have opened up new ways
of exploring two-dimensional materials for thermoelectric
applications\cite{Wei,Zuev,Kato,Buscema,Konabe}.
Since the realization of graphene \cite{Geim,Pesin} there have been numerous 
experimental and theoretical studies
of quasi-2D materials supporting Dirac cones such as silicene \cite{Feng},
germanene \cite{Quhe}, and MoS${}_2$ \cite{Mak}.
Recently, there has been immense interest in synthesis of 2D crystalline
boron structures, generally known as borophene \cite{Mannix,Xu,Zhou}. 
One of them is 8-$Pmmn$ borophene, which is
is a zero gap semiconductor with tilted anisotropic Dirac cones 
\cite{lopez,Zabolotskiy,borophane} and can be thought
of as topologically equivalent to quinoid graphene
\cite{Goerbig1,Goerbig2}.
The bulk optical \cite{Sonu}, magnetotransport \cite{Firoz}, collective modes \cite{Jafari},
Floquet states \cite{Naumis} and thermoelectric properties 
\cite{Zare2} of this borophene structure have been studied extensively.
Another experimentally synthesized allotrope of boron is $\beta_{12}$ 
borophene whose band structure and electronic properties have been extensively 
studied \cite{feng,ezawa}.
Besides, a very recent success of integrating dissimilar
two-dimensional (2D) materials \cite{Xialong} which is essential
for nano-electronic applications, has opened a new direction for studying
the thermoelectricity in junction devices of different materials.
In Ref. \cite{Xialong}, the authors have reported the covalent lateral
stitching of borophene-graphene, resulting in rare realization of 2D lateral
hetero-structure where the lateral interfaces are atomically sharp
despite imperfect crystallographic lattice and symmetry matching.
Furthermore, a graphene/quinoid graphene/graphene junction can be realized
by taking a single graphene sheet, where the middle region is deformed (quinoid graphene) 
by applying a uniaxial strain.

The concept of valleytronics \cite{valley-calori,valley-calori1,valley-calori2,valley-calori3}, 
similar to spintronics 
\cite{Uchida,spin-calo1,spin-calo2,Gerrit, Hatami,Chen}, is becoming popular
in recent past.
In valleytronic devices, the information is
carried by the valley degree of freedom of the charge carriers.
The generation of valley polarization and optically excited
valley-polarized current in various materials have been studied theoretically 
as well as experimentally \cite{valley-pola,valley-pola1,valley-current,valley-current1}.
The harnessing of internal degrees of freedom like spin/valley of the charge carriers
by applying thermal gradient and the associated phenomena are called
spin/valley caloritronics.

Motivated by the above discussion, we study the thermoelectric
effects of a nano-junction system where the left and right electrodes
are made of graphene and the middle region is made of
2D Dirac material having tilted anisotropic Dirac cones, such as borophene
or quinoid graphene. The middle region has different onsite energies on the 
two sublattices and is subjected to circularly polarized electromagnetic radiation. 
It results in valley dependent bands at the two Dirac points and hence a valley 
dependent transmission probability.
Thus, in analogy with the spin-caloritronic studies, thermally activated
quantum transport of valley degree of freedom of the charge
carriers can be achieved.
 Our goal is to study thermally driven valley polarized
properties, known as valley caloritronics and compare them
with that of the charge caloritronics in detail.

This paper is organized as follows. In Sec. (\ref{sec2}),  we present
basic information
of the lateral junction (Subsec. (\ref{seca})) and the definitions of different 
thermoelectric coefficients (Subsec. (\ref{secb})). 
All the numerical results and their corresponding discussions are presented 
in Sec. (\ref{sec3}). Finally, we conclude and summarize our main results 
in Sec. (\ref{sec4}).

\section{Model and Theoretical Methods} \label{sec2}
Here, we first present the essential information of the  junction
characterized by the quasi-ballistic transport.
Later we present general description of 
Seebeck coefficient, electrical conductance, 
thermal conductance and figure of merit for any junction device. 
The discussions in Sec. (\ref{secb}) are
applicable to any junction characterized by the quasi-ballistic transport.

\subsection{Basic information of the junction} 
\label{seca}
\begin{figure}
\begin{center}
\includegraphics[width=92mm]{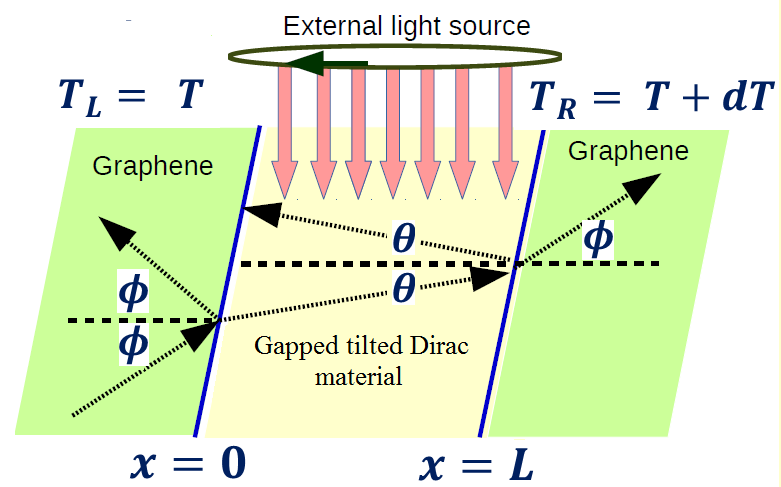}
\caption{
Schematic illustration of the reflection and the transmission processes across
a two-dimensional lateral junction of Dirac material.
The middle region of the junction is illuminated by the circularly polarized 
electromagnetic radiation for opening of
valley-dependent gaps. 
 The angles $\phi$ and $\theta$ denote the incident angles at 
the first and the second interfaces respectively.}
\label{FigSc}
\end{center}
\end{figure}

We consider a two-dimensional junction system placed on the $xy$ plane at room temperature as
shown in  Fig. (\ref{FigSc}). 
The left ($x < 0$) and right ($x > L$) leads are made of graphene sheets, 
while the middle region ($0 < x < L$) is made of a 2D material hosting tilted 
anisotropic Dirac dispersion (can be considered as
borophene or quinoid graphene) with a mass gap at low energy.
Further, it is assumed that the middle region is 
subjected to a circularly polarized electromagnetic radiation where
the photon energy
satisfies the off-resonant condition, i.e., the photon energy is much higher than the band width
of the undriven lattice in the middle region of the system.
The off-resonant circularly polarized  light induces a gap in the energy dispersion. 

The Hamiltonian for the charge carriers in graphene sheet in the vicinity of the 
Dirac points  is given by \cite{Goerbig1}
\begin{equation}
H_G = \zeta\hbar v_F(\sigma_x k_x + \zeta \sigma_y k_y),
\label{eqHg}
\end{equation}
where $\zeta = \pm $ denotes two independent Dirac points, 
$v_F = 10^6$ m/s is the Fermi velocity, $\sigma_x$, $\sigma_y$ are the Pauli matrices 
denoting the sublattice degrees of freedom. The corresponding energy dispersion 
of the Hamiltonian in Eq. (\ref{eqHg}) is given by
$ E_{\lambda}(k) = \lambda \hbar v_F k$, independent of the valley pseudo-spin $\zeta$,
where $\lambda = \pm $ denotes the conduction and valance bands, respectively.
The corresponding eigenfunctions are given by
\begin{equation}
\Psi^{\lambda,\zeta}_G({\bf r}) = \frac{e^{i {\bf k} \cdot {\bf r} }}{\sqrt{2} } 
\left(\begin{array}{c}
1\\
\zeta \lambda e^{i \zeta \phi}
\end{array}\right),
\end{equation}
where $\phi = \tan^{-1}k_y/k_x$.
 
The effective Floquet Hamiltonian, describing the charge carriers of the middle region 
(tilted anisotropic borophene or quinoid graphene) subjected
to circularly polarized electromagnetic radiation, in the vicinity of 
the two independent Dirac points can be written as \cite{Sengupta} (see Appendix A),
\begin{equation}
H_{B} = \zeta\hbar[ v_x q_x \sigma_x + \zeta v_y q_y \sigma_y 
+ v_t \sigma_0 q_y] +  \Delta_\zeta \sigma_z.
\label{eqHqf}
\end{equation}
Here, $\sigma_{0}$ is the $2 \times 2$ identity matrix and  
$ \Delta_\zeta = M + \zeta \Delta $ is the net valley-dependent mass
resulting from the valley-dependent photoinduced mass $\zeta \Delta$ \cite{Kitagawa}
and different onsite energies on the two sublattices $\pm M$  \cite{Haldane}, 
with $M$ being the Semenoff mass.
The photoinduced mass $\Delta = (eA_0)^2 v_x v_y/(\hbar \omega)$ is proportional to the 
intensity of light, which can be tuned experimentally. A tunable Semenoff mass has been 
experimentally achieved in graphene by placing it appropriately on hexagonal boron nitride
 substrate \cite{Jung} or by applying an electric field normal to its plane 
\cite{C1} (which breaks inversion symmetry). Using similar techniques, the creation of 
such a mass gap may be possible in borophene as well, although its experimental realization 
is still not known. We define a dimensionless 
parameter $\eta=M/\Delta$ such that $ \Delta_\zeta = \Delta(\eta + \zeta) $. 
Here, $(v_x,v_y,v_t)$ are the direction dependent 
velocities where $v_t$ (tilt parameter)  is responsible for the tilt in energy dispersion.   
The values of these velocities for borophene
are $v_x = 0.86v_F$, $v_y = 0.69v_F$ and $v_t = 0.32v_F$ \cite{Zabolotskiy}. 
The energy dispersion and the corresponding wave
functions associated with the Hamiltonian in Eq. (\ref{eqHqf}) are given by
\begin{equation} \label{borophene-spec}
E_{\lambda,\zeta}({\bf q}) = \zeta \hbar q v_t \sin\theta + 
\lambda \sqrt{\Delta_\zeta^2 + [\hbar q \Lambda (\theta)] ^2},
\end{equation} 
and
\begin{equation}
\Psi_{B}^{\lambda,\zeta}({\bf r}) = \frac{e^{i {\bf q} \cdot {\bf r} }}{\sqrt{2}} 
\left(\begin{array}{c}
1   \\
\frac{\zeta \hbar q \Lambda(\theta)
e^{i\zeta \delta } }{\Delta_\zeta +
\lambda \sqrt{\Delta_\zeta^2 + [\hbar q \Lambda(\theta)]^2}}   
\end{array} \right),
\end{equation}
where $\delta = \tan^{-1}(v_y q_y/v_x q_x) = \tan^{-1}(\delta_a \tan \theta)$
with $\delta_a=v_y/v_x$, $ \theta = \tan^{-1}(q_y/q_x)$ 
and 
$\Lambda(\theta) = \sqrt{v_{x}^2 \cos^2 \theta + 
v_{y}^2 \sin^2\theta }$ having dimension of velocity.
\begin{figure}
\begin{center}
\includegraphics[width=92mm]{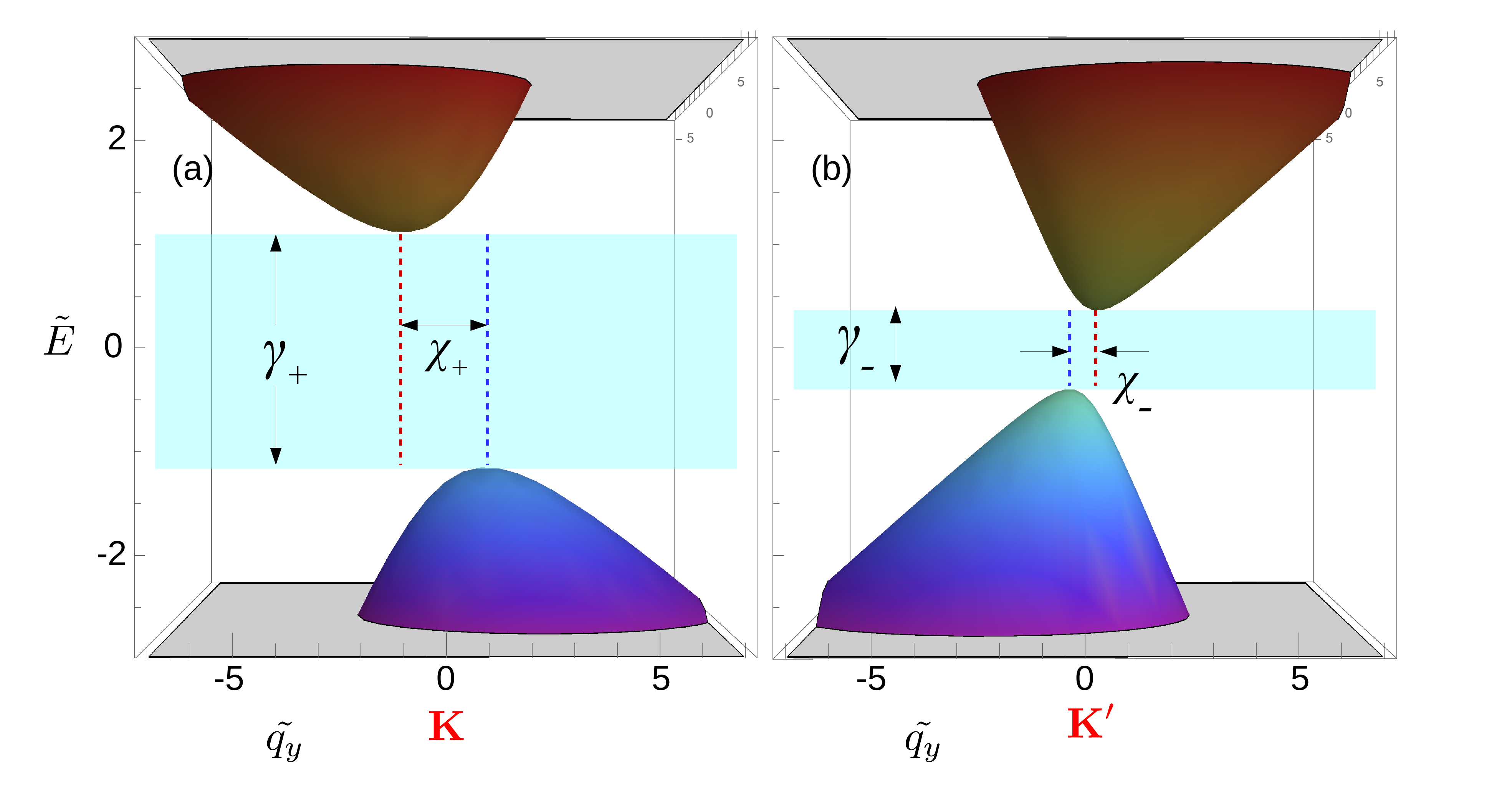}
\includegraphics[width=50mm]{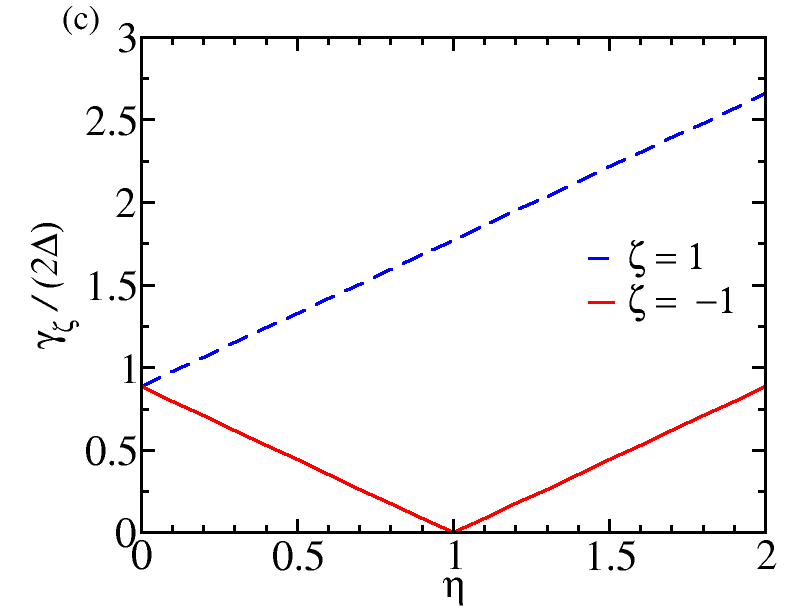}
\caption{(a), (b) Band structure of borophene around the two Dirac points for $\eta=0.5$. Here, 
$\gamma_\zeta$ is the magnitude of the indirect band gap and $\chi_\zeta$ is the 
corresponding shift. The dimensionless variables $\tilde{E}$ and $\tilde{q_y}$ are 
defined as $\tilde{E}=E/\Delta$ and $\tilde{q_y}=\hbar v_F q_y/\Delta$, where $\Delta$ 
is the photoinduced mass. 
$(c)$ The indirect band gaps $\gamma_\zeta$ in units of $2\Delta$ as a 
function of $\eta$.}
\label{Figbands}
\end{center}
\end{figure}

The band structure of borophene in the two valleys with valley-dependent masses 
is shown in Fig. (\ref{Figbands}) for  $\eta=0.5$. The system is an insulator 
with valley-dependent indirect band gaps $\gamma_\zeta$. The shift $\chi_\zeta$ 
between maxima of valence band and minima of conduction band in both the valleys 
is along the $q_y$ axis. The magnitude of the indirect gaps and the shifts are given as
\begin{equation}
\label{eqbb}
\gamma_\zeta=2\Delta_\zeta\sqrt{1-\frac{v_t^2}{v_y^2}}, \hspace{0.5cm} 
\chi_\zeta=\frac{2 v_t \Delta_\zeta}{\hbar v_y \sqrt{v_y^2-v_t^2}}.
\end{equation}
Equation (\ref{eqbb}) reveals that the band gaps reduce due to tilt and 
decrease monotonically with $v_t$ for $v_t<v_y$ while the shifts corresponding to the gaps increase. 
For $\eta > 0$, the band gap at  $K^\prime$ valley is smaller than at  $K$ valley. 
So, the effective band gap of the system is $\gamma_-$. The direct gaps at the original 
Dirac points are equal to $2\Delta_\zeta$.  The magnitude of the gap at 
$K$ valley monotonically increases with  $\eta$, whereas for  $K^\prime$ valley, 
it initially decreases with $\eta$ (for $\eta<1$), vanishes at $\eta=1$ and 
then starts to increase again for $\eta>1$ (see Fig. (\ref{Figbands}c)).

The middle region can be reduced to a gapped
graphene by setting $ v_t = 0$ and $v_x = v_y = v_F$, so that the junction becomes 
a graphene/gapped-graphene/graphene junction.
If the Fermi energy lies in the gap, the middle region behaves like a 
topological insulator when $M < \Delta$, otherwise it is a trivial insulator. 
In the topological insulating state, the edge states contribute to the transport quantities.
Since our system is kept at room temperature, the contribution from the bulk states 
would dominate over the edge states contribution \cite{Murakami}.

Suppose an electron from the left lead is injected with an energy $\epsilon$ and 
incident angle $\phi$.
The valley-dependent transmission probability 
$ \mathcal T_\zeta(\epsilon,\phi) = |t_\zeta(\epsilon,\phi)|^2$ of the electron from left to right lead  
is obtained as (see appendix \ref{ATP})
\begin{widetext}
\begin{equation}
\label{eqT}
\mathcal T_\zeta(\epsilon,\phi) = \frac{4p_{\zeta}^2 \cos^2\phi \; (1+\cos2 \delta)}
{(2\sqrt{2} p_{\zeta} \cos \phi \cos\delta)^2 + 
[1 + p_{\zeta}^2 - 2p_{\zeta}\cos(\phi-\delta)]
[1 + p_{\zeta}^2 + 2p_{\zeta}\cos(\phi + \delta)][1- \cos(2q L  \cos\theta)] }.
\end{equation}
\end{widetext}
where $p_{\zeta}$ is given by
\begin{eqnarray}
\label{eqp}
p_{\zeta} & = & 
\frac{\hbar q \Lambda(\theta)  }
{\Delta_\zeta + \sqrt{\Delta_\zeta^2 + [\hbar q \Lambda(\theta)]^2}}.
\end{eqnarray}
The values of $q$ and $\theta$ in the expression of $p_{\zeta}$ can be obtained 
by solving the following two coupled 
equations: 
\begin{eqnarray}
\label{eq11}
 q \sin \theta &= & k \sin \phi, \\\nonumber
q & = & \frac{\zeta \epsilon v_t \sin \theta \mp 
\sqrt{\Lambda^2(\theta)(\epsilon^2 -\Delta_\zeta^2) + (v_t \Delta_\zeta \sin \theta)^2} }
{\hbar[(v_t \sin\theta)^2 - \Lambda^2(\theta)]}.
\end{eqnarray}

\begin{figure}
\begin{center}
\includegraphics[width=89.5mm, height=45mm]{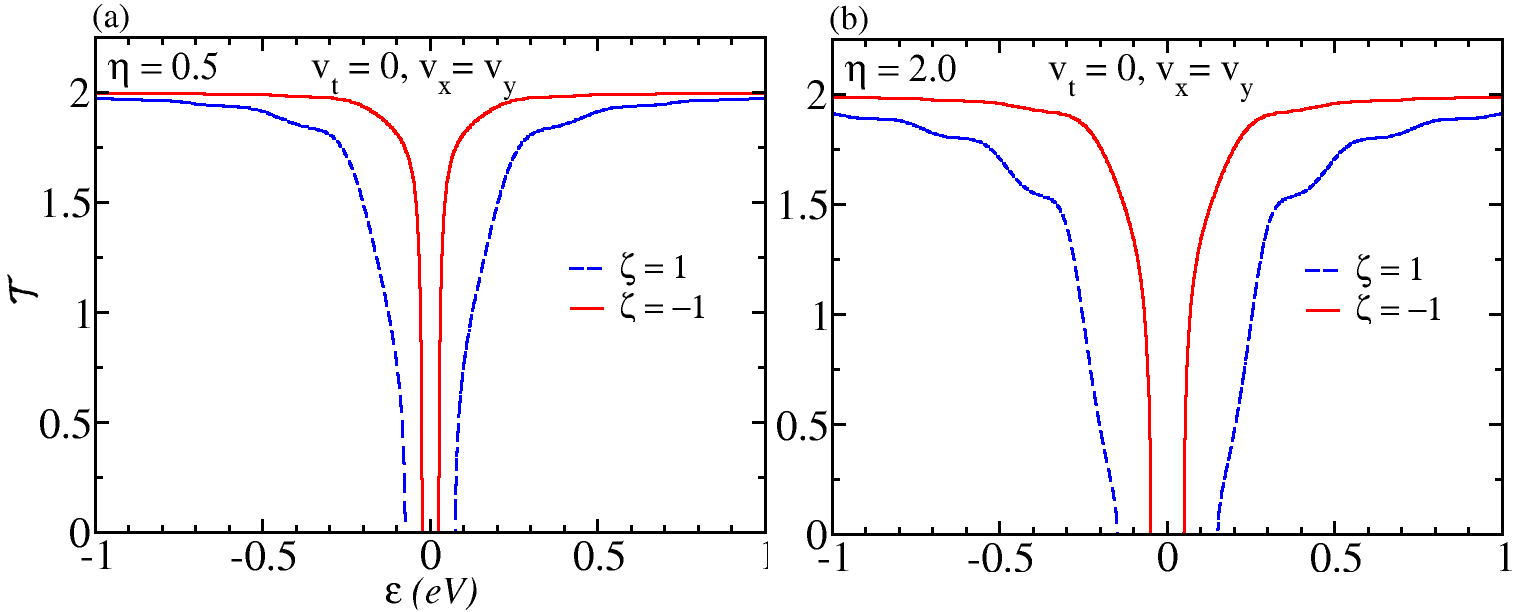}
\caption{ $\mathcal T_\zeta(\epsilon)$ vs $\epsilon$ for a) $\eta = 0.5$, b) $\eta = 2 $ when $v_t = 0$, 
$v_x = v_y=v_F$.}
\label{FigTEG}
\end{center}
\end{figure}
\begin{figure}
\begin{center}
\includegraphics[width=89.5mm, height=45mm]{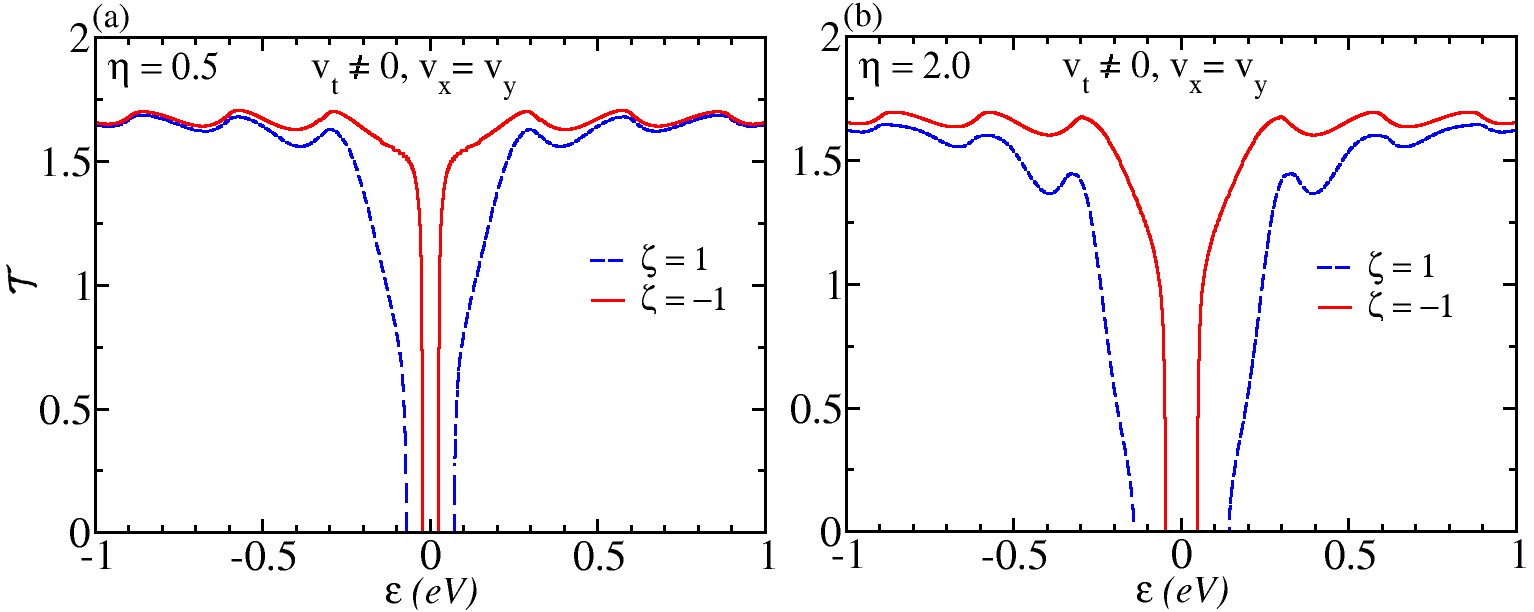}
\caption{$\mathcal T_\zeta(\epsilon)$ vs $\epsilon$ for a) $\eta = 0.5$, b) $\eta = 2 $ when $v_t\neq 0$, 
$v_x = v_y=v_F$.}
\label{FigTEB}
\end{center}
\end{figure}
We define the effective  
transmission coefficient for current along $x$ direction at a given energy $\epsilon$ as 
$\mathcal T_\zeta(\epsilon) = \int_{-\pi/2}^{\pi/2} \mathcal T_\zeta(\epsilon,\phi)\cos\phi d\phi $. 
In Figs. (\ref{FigTEG}), (\ref{FigTEB}),  $\mathcal T_\zeta(\epsilon)$ vs. $\epsilon$ is plotted 
for two conditions -- 
$(i)$ $v_t=0$, $v_x=v_y=v_F$ and $(ii)$  $v_t\neq0$, $v_x=v_y=v_F$. 
The oscillations in $\mathcal T_\zeta(\epsilon)$ in Fig. (\ref{FigTEB}) appear 
due to the $\cos (2qL\cos\theta)$ term (see Eq. (\ref{eqT})), where $q$ is a  
function of $\epsilon$. 
In  case of $v_x=v_y=v_F$, $v_t=0$  
and $E \gg \Delta_{\zeta}$, the values of the  $\phi$ and $\delta$ become  
equal and $p_{\zeta}$ becomes $\sim 1 $, which eventually yields the coefficient 
of $(1-\cos (2qL\cos\theta))$ to be $\sim 0$ and thus $\mathcal T_{\zeta}(\epsilon,\phi)$ becomes 
$\sim 1$ (see Eq. (\ref{eqT})) and $\mathcal T_\zeta(\epsilon)\sim 2$.  
As a result no such noticeable oscillations are obtained 
(see red curves of Figs. (\ref{FigTEG}a) and (\ref{FigTEG}b)). 
The physics behind this can be explained using the concept of electron wave 
interference -- when $v_t=0$, $v_x=v_y=v_F$ and 
$E \gg \Delta_{\zeta}$, the system can be viewed as a single graphene sheet without any 
barrier. Thus, almost all the incoming electron waves from the left lead get transmitted to 
the right lead, leaving almost no reflected wave and thereby causing no
interference. If the band gap is further increased, the probability of the reflection 
of electron waves from the interface increases; so the reflected and the transmitted 
electron waves begin to interfere. This results in oscillations in the transmission 
probability for $v_t=0$, $v_x=v_y=v_F$ case as well (see blue curves in Figs. (\ref{FigTEG}a) and (\ref{FigTEG}b)).
Furthermore, the $K$ valley has smaller $\mathcal T_\zeta(\epsilon)$ as compared to $K^\prime$ valley.
It can be understood using the analogy of transmission through a rectangular potential barrier. 
If the middle region is considered as a potential barrier with a barrier height $\Delta_{\zeta}$, 
the transmission probability is smaller for a larger barrier height for the considered range of incident 
energies above the barrier. Since $\Delta_{+}>\Delta_{-}$, $K$ valley allows lesser transmission than $K^\prime$.  

The tilted velocity term diminishes $\mathcal T_\zeta(\epsilon)$, as for $v_t\neq 0$ 
the transmission probability as a function of incident angle  shows  more deviation from 
$1$ compared to $v_t=0$ case (see Figs. (\ref{figTNG}), (\ref{figTG}) and (\ref{figTB})). 
Moreover, the $\mathcal T_\zeta(\epsilon)$ is almost electron-hole symmetric 
(see Figs.  (\ref{FigTEG}), (\ref{FigTEB})) although $v_t$ 
breaks the electron hole symmetry in band structure (see Fig. (\ref{Figbands})).

\subsection{Thermoelectric coefficients}
\label{secb} 

A detailed derivation of the thermoelectric coefficients is given in appendix (\ref{ATTE}).
The valley resolved Seebeck coefficient $S_{\zeta}$ for a small temperature difference $dT$ is given as 
\begin{equation}
\label{EqnS}
S_{\zeta} = -\frac{dV_\zeta}{dT}\Big|_{dI_{\zeta}=0} = - \frac{L^{(1)}_{\zeta}}{eTL^{(0)}_{\zeta}}.
\end{equation}
where $dV_\zeta$ are the valley-resolved thermoemfs induced between the cold 
and hot leads and  $L^{(\alpha)}_{\zeta}$ are the kinetic coefficients for quasi-ballistic transport 
given by 
\begin{eqnarray}
\label{eqnlz}
L^{(\alpha)}_{\zeta}& =& \int_{-\pi/2}^{\pi/2}d\phi \cos\phi 
\int_{-\infty}^{\infty} \mathcal T_{\zeta}(\epsilon,\phi )N(\epsilon)(\epsilon - \mu)^{\alpha} \\ \nonumber
& \times &
\big(-\frac{\partial f}{\partial \epsilon} \big) d\epsilon 
\end{eqnarray}
with $\alpha = 0,1,2$.

At zero external bias voltage $(V_B=0)$, the valley resolved electrical conductance
$G_{\zeta}$ can be expressed as
\begin{equation}
G_{\zeta} = \frac{dI_{\zeta}}{dV}\Big|_{V_B=0} = \frac{2e^2}{h}L^{(0)}_{\zeta}
\end{equation}
where $I_{\zeta}$ is the valley-dependent 
charge current  given in  Eq. (\ref{elec-current}).

 The valley resolved thermal conductance $k^{\rm el}_{\zeta}$ associated with 
the valley-dependent thermal currents $J^{\rm el}_\zeta$ can be expressed 
in terms of the kinetic coefficients $L^{(\alpha)}_{\zeta}$ (see Eq. (\ref{eqnlz})) 
as \cite{Mawrie,Liu} 
\begin{equation}
\label{eq48}
k^{\rm el}_{\zeta} = \frac{2}{h}\frac{L^{(2)}_{\zeta}}{T} +  \frac{2e}{h}L^{(1)}_{\zeta}S_{\zeta}.
\end{equation}

The total charge and thermal conductance are defined as $G_c = G_+ + G_-$ and 
$k^{\rm el}_c = k^{\rm el}_+ + k^{\rm el}_-$.  Similar to Ref. \cite{Gerrit, Hatami}, 
we define the charge Seebeck 
coefficient as $S_c = (S_+G_+ + S_-G_-)/(G_+ + G_-)$.  The charge Seebeck coefficient can be viewed 
as the effective thermoemf generated between the two leads per unit temperature difference.

In this system, the carriers in the two valleys have unequal transmission probabilities owing to 
distinctive nature of the valley gaps. So, the heat and particle flux in the two valleys differ, 
giving rise to valley polarized charge currents $ (I_v = |I_+ - I_-|)$ and thermal currents 
$ (J^{\rm el}_v = |J^{\rm el}_+ - J^{\rm el}_-|)$. This leads to different induced voltages 
in the two valleys at the cold lead. Thus, the two valleys act as two conducting channels having
different thermopowers present within the same system. Since the transfer of electrons between 
the valleys is prohibited due to large separation of the valleys in (quasi)momentum space and 
absence of any valley-mixing mechanism, a valley emf ($dV_+-dV_-$) exists. This phenomenon can be 
termed as the valley Seebeck effect analogous to the spin Seebeck effect \cite{Uchida}. 
It refers to the generation of a valley voltage resulting from a temperature gradient. 

Using the same analogy as in Ref. \cite{Chen} for spin Seebeck coefficient, the valley Seebeck coefficient 
is defined as $S_v = |S_+ - S_-|$. Similar to the spin Seebeck effect \cite{Uchida}, 
the valley Seebeck coefficient can be viewed as the potential difference between charge 
carriers of the two valleys in the cold lead per unit temperature difference. 
Similar to the valley current, the valley polarized electrical conductance and 
the valley polarized thermal conductance can be defined as
$ G_v = |G_+ - G_- |$ and $k^{\rm el}_v=|k^{\rm el}_+ - k^{\rm el}_-|$ respectively.

One of the challenges in fabricating thermoelectric devices
is to obtain optimal conditions which ensure the operation
of the device with maximum power output at the best possible efficiency. 
The efficiency of the system depends upon a quantity called
the figure of merit $Z_cT$ which is defined as
\begin{equation}
\label{eqch5FM}
Z_cT = \frac{S_c^{2}G_c}{k^{\rm el}_c + k^{\rm ph}}T,
\end{equation}
where $S_c$ is charge Seebeck coefficient, $G_c$ is the charge conductance, $k^{\rm el}_c$ is 
the thermal conductance of the carriers, $k_{\rm ph}$ is phonon's thermal conductance  
owing to the involvement of lattice structure and $T$ is the absolute temperature. 
 The possibility of extracting the valley thermoemf for power generation allows 
us to define valley figure of merit $Z_v T = \frac{S_v^2 G_v}{k^{\rm el}_c}$ of this device using 
the same analogy as spin figure of merit in Ref. \cite{Swirkowicz,Chen,Wierzbicki,Rameshti}. 

In the context of phononic contribution to thermal conductance, we would like to mention 
that the Debye temperature in borophene has a high value about $2000 $ K \cite{Tohei,Z} 
due to the strong bonding. Further, graphene 
has also a higher Debye temperature $\theta_D = 2300 $ K, approximately an order of magnitude 
higher than for typical metals.
Thus, the room temperature ($300 $ K) is safely assumed
to be low with respect to high Debye temperatures of borophene and graphene.  
Due to this reason, the phonon population is expected to be low at room temperature, 
which diminishes the possibility of phonon-phonon inelastic scattering 
events. Henceforth we neglect the phonon's contribution in thermal conductance.

\section{Results and Discussion} \label{sec3}
Here we present numerical results of different thermoelectric properties 
of the junction subjected to the off-resonant Floquet radiation.
For our numerical analysis, we choose the parameters
$v_x =v_y = v_F, v_t = 0.32v_F$ and $\Delta = 0.05$ eV.
The dimensionless parameter, $\eta=M/\Delta$  is varied in the range [$0:1.5$].
 It should be noted that with increasing $\eta$ value further, the conductance 
in $K$ valley vanishes in the low chemical potential region. As our main goal is to study 
the valley polarized properties, to get non-zero values of the conductance for both the valleys, 
we choose the maximum $\eta=1.5$.     
Temperature of the cold lead is maintained at $T = 300$ K and that of the 
hot lead is $T+dT$ where $dT\ll T$. The dimensions of system are taken as $(L, W ) = (50, 30)$ nm.

\subsection{ VALLEY ELECTRICAL CONDUCTANCE}

The variation of $G_v$ and $G_c$ with chemical potential $\mu$ for different values 
of $\eta$ are shown in Fig. (\ref{figG}a) and  Fig. (\ref{figG}b) respectively.

{\bf (a) Dependence on chemical potential}: 
\begin{figure}
\begin{center}
\includegraphics[width=88.5mm, height=45mm]{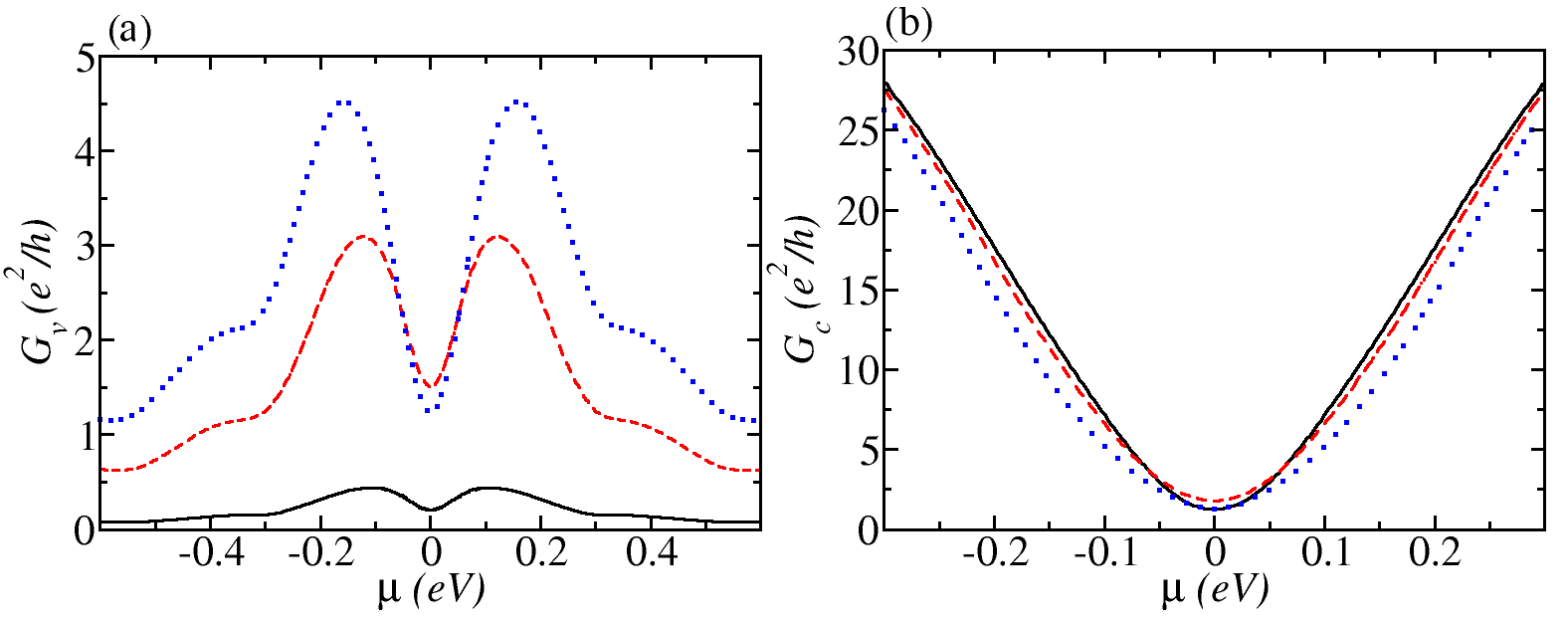}
\caption{Variation of (a) valley conductance $G_v$ and (b) charge conductance $G_c$ as a 
function of chemical potential $\mu$ for different values of $\eta$: 
$\eta = 0.1$ (black solid), $\eta=0.8$ (red dashed) and $\eta=1.5$ (blue dotted).}
\label{figG}
\end{center}
\end{figure}

The valley polarized conductance $G_v$ has peaks at $\pm\mu_p(\eta)$ and a local minimum 
at $\mu=0$ (Fig. \ref{figG}a). This feature is also present when the middle region is gapped 
graphene with unequal masses $\Delta_\zeta$ in the two valleys, which 
indicates that tilt is not responsible for the peaks. The appearance of peaks 
can be explained using an analogy with transmission through a rectangular barrier. 
For gapped graphene, the dispersion can be approximated as 
$E_\zeta \approx \Delta_\zeta + \frac{\hbar^2 q^2}{2 (\Delta_\zeta/v_F^2)}$ near the 
band minima/maxima. So, the middle region can be viewed as a potential barrier with 
valley-dependent barrier heights ($V_\zeta=\Delta_\zeta$) and effective masses 
$(m_\zeta=\Delta_\zeta/v_F^2)$. The rate of increase of transmission 
$\mathcal T(\epsilon)$ with $|\epsilon|$ for smaller mass ($\Delta_-$) is higher 
than that with larger mass ($\Delta_-$) for energies just above the barrier 
(see Figs. (\ref{FigTEG}) and (\ref{FigTEB})). Since $G_\zeta(\epsilon)$ is 
proportional to $\mathcal{T}_\zeta(\epsilon)$, $G_+$ increases slowly  with $\mu$ resembling 
a quadratic growth, while $G_-$ rises sharply resembling almost a linear growth due 
to smaller mass. For higher value of $|\mu|$, the effect of mass in the dispersion becomes 
negligible in both the valleys which results in almost similar variation of their conductances 
with $|\mu|$. Due to this nature, $G_v$
increases with $|\mu|$ initially, attains a maximum (peak) and then decreases asymptotically 
to zero at higher $|\mu|$.   

On the other hand, $G_c$ increases monotonically with increase in $|\mu|$ (Fig. \ref{figG}b). 
This is primarily due to the increase in the number of available conducting channels 
$N(\epsilon)$ in the leads with increase in $|\epsilon|$. As expected, $G_c$ is always 
greater than $G_v$ for a given $\eta$.
 
{\bf (b) Dependence on gap parameter}:
The valley conductance $G_v$ increases with the increasing strength of $\eta$ for $\eta<1$ and starts to decrease  with $\eta$ for $\eta>1$. 
 This can be explained as follows -- Since a larger gap 
corresponds to lesser $\mathcal T_\zeta(\epsilon)$, the increase in $\eta$ lowers $G_+$, owing 
to the monotonic rise in $\Delta_+$ with $\eta$ (see Fig. \ref{Figbands}(c)). Similarly, due to the non-monotonic 
variation of $\Delta_-$, $G_{-}$ initially increases with $\eta$, attains maximum value 
at $\eta=1$ and then starts to decrease with $\eta$. Since $G_-$ increases while $G_+$ decreases with $\eta$ for $\eta<1$, their difference 
gets enhanced with $\eta$. For $\eta>1$, both $G_{+}$ and $G_{-}$ decrease with $\eta$, but the rate of decrease of $G_{-}$ is more than that of $G_+$. As a result, $G_v$ gets diminished with $\eta$ for $\eta>1$. For $\eta\to 0$, we get $\Delta_{\zeta}\to  \zeta\Delta$ which 
yields $\mathcal T_{+}(\epsilon,\phi)=\mathcal T_{-}(\epsilon,-\phi)$.  
On integrating out $\phi$, both the valleys give the same value of transmission at a 
given energy. Hence, $G_v \to 0$ as $\eta\to 0$. 

Figure \ref{figG}(b) reveals that $G_c$ gets diminished (though the change is small, but noticeable) 
with increasing $\eta$, away from the low chemical potential regime. Near the low chemical 
potential region, there are crossovers in the conductance plots.\\

The tilt $v_t$ diminishes $G_\zeta$ in each valley, which results in lowering of $G_c$. 
It is found that $G_v$ also decreases with $v_t$, since  $v_t$ diminishes 
$ (\mathcal{T}_-(\epsilon) - \mathcal {T}_+(\epsilon))$ as well 
(see Figs. (\ref{FigTEG}) and (\ref{FigTEB})). It is interesting to note that charge and 
valley conductances show high degree of electron-hole symmetry despite the fact that 
the spectrum in the middle region is electron-hole asymmetric due to non-zero $v_t$. 
Similar behaviour is shown in bulk borophene \cite{Zare2}.

\subsection{ VALLEY SEEBECK COEFFICIENT (THERMOPOWER)}\label{seebecksec}

\begin{figure}
\begin{center}
\includegraphics[width=88.5mm, height=45mm]{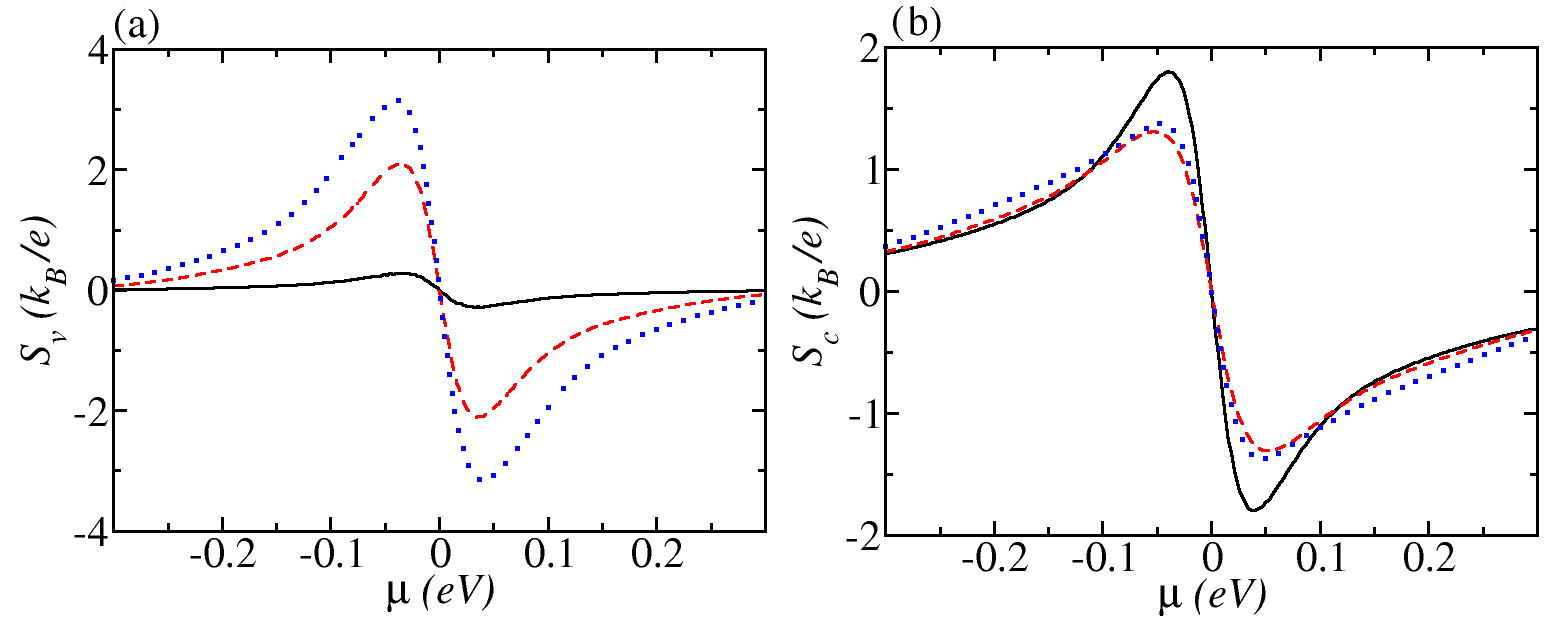}
\includegraphics[width=50mm, height=45mm]{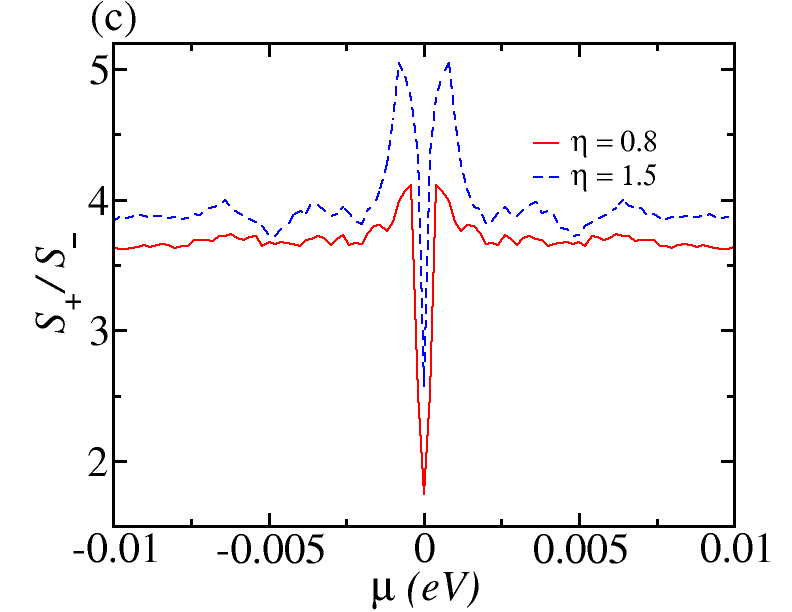}
\caption{ Variation of (a) valley Seebeck coefficient $S_v$, (b) charge Seebeck coefficient 
$S_c$ and   (c) ratio of $S_+$ and $S_-$
as a function of chemical potential $\mu$ for different values of $\eta$:
$\eta=0.1$(black solid), $\eta=0.8$ (red dashed) and $\eta=1.5$ (blue dotted).}
\label{figS}
\end{center}
\end{figure}

The valley and charge Seebeck coefficients $S_v$ and $S_c$ (in units of $k_B /e$) as a 
function of the chemical potential $\mu$ for various values of $\eta$ are shown in 
Fig. (\ref{figS}a) and Fig. (\ref{figS}b) respectively .

{\bf (a) Dependence on chemical potential}:
Both $S_v$ and $S_c$ display a local maxima (minima) at $\mu_s \sim 0.04 \;(-0.04 )$ eV 
on variation with $\mu$. The value of $\mu_s$ is roughly independent of $\eta$. 
The maxima/ minima in the thermopower arises due to the term 
$(\epsilon-\mu)(-\frac{\partial f}{\partial \epsilon})$ in the numerator of Seebeck 
coefficient (see Eq. (\ref{EqnS})).  There is a change in sign of $S$ while change in 
sign of $\mu$ (due to $(\epsilon - \mu)$ term in the numerator of $S$). It indicates the 
change in electrical nature of the charge carriers as $\mu$ changes sign. 
When $\mu$ lies in the conduction (valence) band, thermally activated electrons (holes) 
propagate opposite (parallel) to the temperature gradient, which results in negative (positive) 
thermopower. Similar to the conductance, the electron-hole symmetry is nearly perfect in the 
absolute value of Seebeck coefficients. 

{\bf (b) Dependence on gap parameter}:
The absolute values of $S_v$ increase with $\eta$ at a given chemical potential. 
This can be explained as follows -- From the definition of Seebeck coefficient 
$ S_{\zeta} \sim \frac{L^{(1)}_{\zeta}}{G_{\zeta}}$, we can say that increase in 
$L^{(1)}_{\zeta}$ and decrease in $G_{\zeta}$ with $\eta$ aid to 
enhance the value of $S_\zeta$.  Since $L^{(1)}_+$ increases and $G_{+}$ decreases 
with $\eta$, $S_+$ starts to gain weight as we increase $\eta$. Similarly, $S_-$ 
initially decreases with $\eta$, attains minimum at $\eta=1$ and again starts to increase. 
Though $S_+$ and $S_-$ show different 
nature of variation with $\eta$, their difference as a function of $\eta$ is mainly dictated 
by $S_+$. This happens because 
$K$ valley's contribution in thermopower changes more rapidly with $\eta$ as compared 
to $K^\prime$ valley .  The behavior of valley thermopower as a function of $\eta$ 
can be understood from Fig. (\ref{figS}(c)) also.  As the ratio of $S_+$ and $S_-$ increases 
with $\eta$, the valley thermopower gets enhanced  with increasing strength of $\eta$. 
The reason behind this can be understood from Fig. (\ref{FigTEB}), 
as it reveals that with increasing $\eta$ from $0.5$ to $2.0$, the rate of change 
in $\mathcal T(\epsilon)$ in $K$ valley is more than that in $K^\prime$ valley. 

For the chemical potential $|\mu|<0.08 $ eV, the charge thermopower $S_c$  decreases with 
$\eta$ for $\eta<1$, attains minima around $\eta=1$, and then starts increasing again. 
For $|\mu|>0.08 $ eV, an increase in $\eta$ aids to enhance $S_c$, though the enhancement is quite small. 
So $S_v$ and $S_c$ behave differently with $\eta$ which is mainly due to the different weightages 
of $K$ and $K^{\prime}$ channels' contribution in their definitions. It seems that on increasing the strength 
of $\eta$ even more, one can achieve higher valley thermopower. But it is not possible, as 
for such a higher value of $\eta$, there will be no available channel to conduct 
in the low chemical potential regime.

It is  worth mentioning that $S_v$ decrease with $v_t$. Though $v_t$ diminishes $G_\zeta$ 
(denominator of $S_\zeta$), it lowers $L_{\zeta}^{(1)}$ (numerator of $S_{\zeta}$) too. 
Hence, the contribution of $(\frac{L_{+}^{(1)}}{G_{+}}-\frac{L_{-}^{(1)}}{G_{-}})$ 
(see definition of $S_v$) decreases as we increase $v_t$. The charge thermopower also
behaves similarly as a function of $v_t$.

The materials with large electron-hole asymmetry are known to enhance the 
thermoelectric coefficient. So more thermopower is expected when the middle region 
is made of tilted Dirac material instead of graphene. But $v_t$ does not break 
the electron-hole symmetric nature in $\mathcal T_\zeta(\epsilon)$ as shown in 
Fig. (\ref{FigTEB})). Hence, it cannot aid to enhance the thermopower of the system. 
It is to be noted that if the middle region is also monolayer graphene, 
then for higher value of $\eta$, we do not get any thermopower in the low chemical 
potential regime, whereas if the middle region is borophene or quinoid 
graphene, we get finite values of thermopower for those low values of chemical potential. 
The physics behind this as follows -- for graphene, in case of $ \epsilon < |\Delta_{\zeta}|$, 
there is no transmission 
because of imaginary momentum, and thus no channel to conduct. But for borophene or 
quinoid graphene,  due to the indirect gap for the tilted velocity term, the momentum 
is real until $\epsilon > |\gamma_{\zeta}/2|$, and hence few channels are available to 
conduct, although $\epsilon < |\Delta_{\zeta}|$ (See Eq. (\ref{eqbb}) and Fig. (\ref{Figbands})).
Thus for low chemical potential, the highly gapped graphene-borophene-graphene junction 
is good candidate with respect 
to highly gapped graphene-graphene-graphene junction as a thermoelectric device.

\subsection{ VALLEY THERMAL CONDUCTANCE}

\begin{figure}
\begin{center}
\includegraphics[width=90mm, height=46mm]{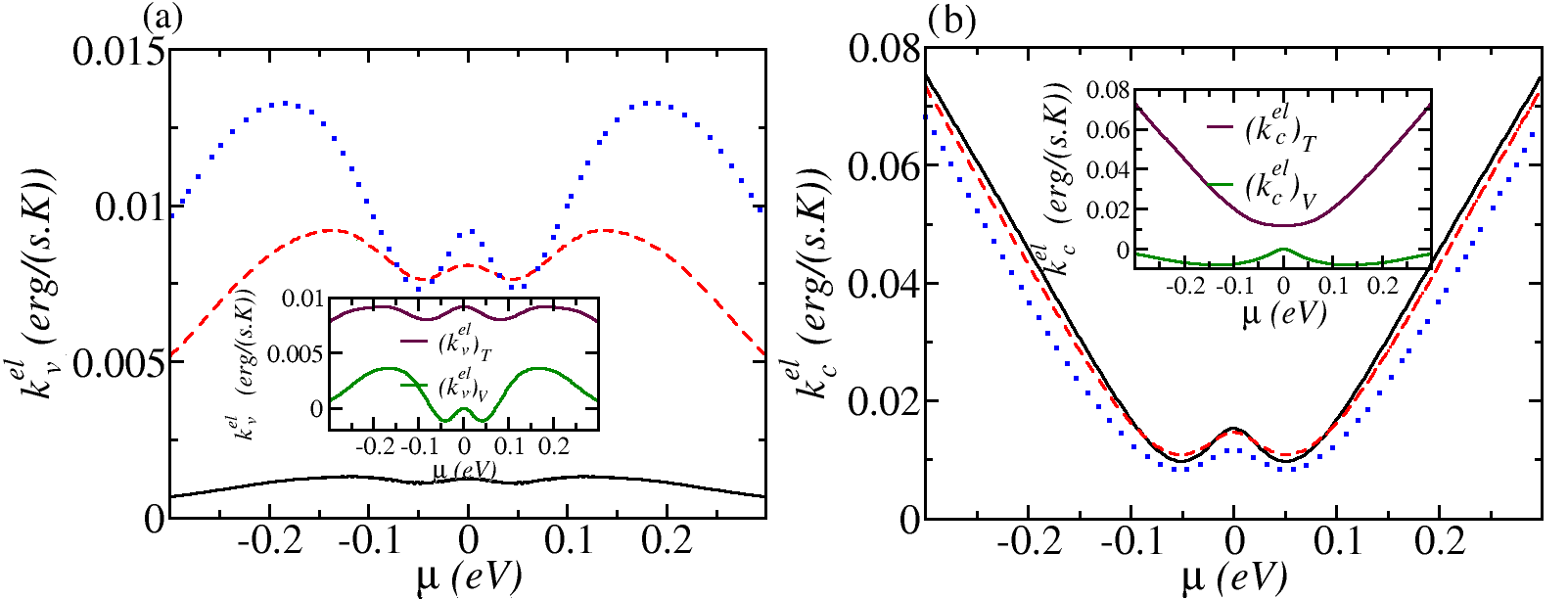}
\caption{Variation of (a) electrical valley thermal conductance $k^{\rm el}_v$ and (b) 
electrical thermal conductance $k^{\rm el}_c$ as a function of 
chemical potential $\mu$ for different values of $\eta$:  $\eta=0.1$ (black solid), 
$\eta=0.8$ (red dashed) and $\eta=1.5$ (blue dotted).}
\label{figK}
\end{center}
\end{figure}

The valley polarized thermal conductance $k^{\rm el}_v$ and charge thermal conductance 
$k^{\rm el}_c $ as a function of $\mu$ are shown in Fig. (\ref{figK}a)  
and Fig. (\ref{figK}b) respectively for different values of $\eta$.

{\bf (a) Dependence on chemical potential}:

As the valley resolved thermal conductance $k_{\zeta}^{\rm el}$ arises due to the energy 
flow carried by the charge carriers, $k^{\rm el}_{v/c}$ as a
function of $\mu$ shows almost similar features as $G_{v/c}$ except 
the region where $\mu$ is close to zero. In the case of $k^{\rm el}_{v/c}$ 
(see Figs. (\ref{figK}a) and (\ref{figK}b) ) 
there is a bump near $\mu=0$, while for $G_{v/c}$ there is no such thing. 
This bump in $k^{\rm el}_{c}$ arises due to the 
$(k^{\rm el})_V$ term as shown in the inset of Fig. (\ref{figK}b), whereas valley $(k^{\rm el})_T$ (in addition to valley $(k^{\rm el})_V$) is also responsible for the bump in $k^{\rm el}_{v}$ (see inset of Fig. (\ref{figK}a)).
Moreover, the valley thermal conductance $k^{\rm el}_v$ has peaks at $\pm\mu_p(\eta)$ and then starts decreasing with $\mu$ (similar to $G_v$), 
as opposed to $k^{\rm el}_c$.   

{\bf (b) Dependence on gap parameter}:
As expected, $k^{\rm el}$ shows the same nature as the electrical charge conductance as a 
function of $\eta$, which is depicted in Fig. (\ref{figK}). The reason behind this nature 
is same as for the charge conductance. 

Not surprisingly, $k^{\rm el}$
as a function of $v_t$ shows the similar behavior as electrical charge conductance, 
indicating that electrical thermal conductance is diminished by $v_t$. 
Here we would like to mention 
that in our system, the Wiedmann-Franz law which states that $\sigma_c/\kappa_c = LT$, where 
$L = 2.44 \times 10^{-8}$ W$\Omega {\rm K}^{-1}$ is 
the Lorentz number, $\sigma_c$ is electrical charge conductivity and $\kappa_c$ is the 
electrical thermal conductivity, holds good for low temperature, though it  deviates
near $\mu=0$.  This law is valid in case of valley polarized conductivities as well under the same conditions.

\subsection{ VALLEY FIGURE OF MERIT}

\begin{figure}
\begin{center}
\includegraphics[width=89.5mm, height=45mm]{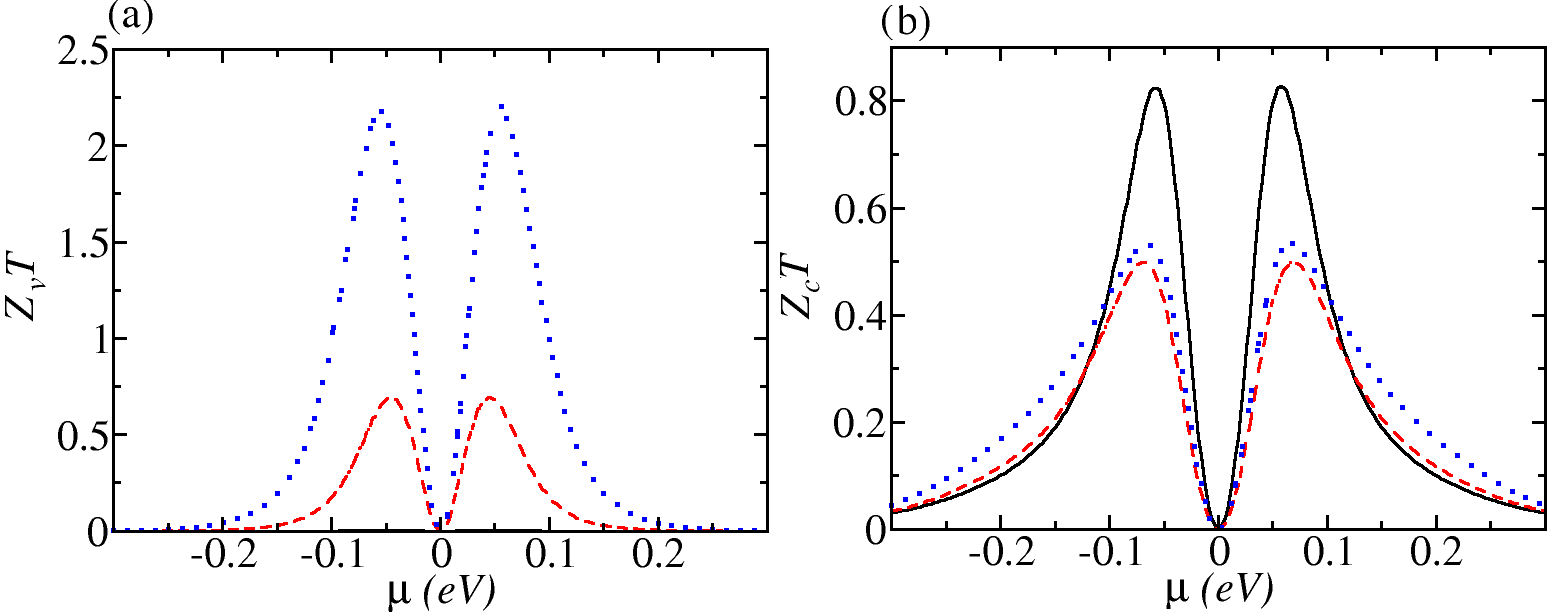}
\caption{ Variation of (a) valley figure of merit $Z_vT$ and (b) charge figure of merit 
$Z_c T$ as a function of 
chemical potential $\mu$ for different values of $\eta$: $\eta=0.1$ (black solid), 
$\eta=0.8$ (red dashed) and  $\eta=1.5$ (blue dotted).}
\label{figZ}
\end{center}
\end{figure}

The variation of the valley and charge figures of merit $Z_vT$ and $Z_cT$ with $\mu$ are shown in 
Fig. (\ref{figZ}a) and Fig. (\ref{figZ}b) respectively for different values of $\eta$.

{\bf (a) Dependence on chemical potential}: Both the figures of merit show similar behavior as a 
function of $\mu$ and have maxima close to $\mu \sim \pm 0.06$ eV. 
The occurrence of maxima can be explained as follows -- From the definition of $ZT$, 
we see that $ZT$ varies as $S^2$. Since $S_v$ and $S_v$ attain extrema near $\mu \sim \pm0.06$ eV 
[see \ref{seebecksec}(a)], $Z_vT$  and $Z_cT$ are also peaked 
around those values  for the given set of parameters. The positions of the peaks are almost 
insensitive to the gap parameter $\eta$.

{\bf (b) Dependence on gap parameter}:
Increase in $\eta$ enhances $Z_vT$, because $Z_vT$ mainly varies as $S_v^2$ which shows increasing trend with $\eta$. 
For $\eta\to 0$, $Z_vT\to 0$ as the valley thermopower $S_v$ vanishes. The charge figure of merit 
behaves non monotonically  with $\eta$ for $|\mu|<0.1 $ eV, while for $|\mu|>0.1 $ eV 
it increases with $\eta$, thereby depicting the trend of $S_c^2$.  For $|\mu|<0.1 $ eV, 
the $Z_cT$ decreases with $\eta$ (for $\eta<1$), becomes minimum at $\eta=0$ and starts to 
increase again for $\eta>1$.

It is observed that both $Z_vT$ and $Z_cT$ get reduced with increase in $v_t$. This is attributed to the fact 
that increase in $v_t$ reduces the thermopower (see \ref{seebecksec}) while the ratio of 
$G$ and $k^{\rm el}$ does not vary appreciably with $v_t$. For the parameters 
used in the problem, the maximum values of $Z_vT$ and $Z_cT$ are 2.2 and 0.82 respectively. 
With the inclusion of phonon's thermal conductance using the value 
in Ref. \cite{Xiao}, $(Z_vT)_{\rm max}$ and $(Z_cT)_{\rm max}$ are reduced to 
$\sim 1.84$ and $\sim 0.67$ respectively. However, the value mentioned in the Ref. \cite{Xiao} 
is for pristine borophene, whereas our system is a hetero-junction with a band gap in 
the dispersion of middle region. Thus the values are not accurate, rather an estimation. 

Here we would like to mention that the structure of borophene is 
anisotripic along $x$ and $y$ directions and it  
affects on the transport properties in two directions differently. For bulk borophene cases 
in Ref. \cite{Zare2}, the electrical and thermal conductance  in two directions differ 
quantitatively rather than qualitatively, while the results obtained for thermopower and 
figure of merit are almost direction independent. Since the results in both directions are 
qualitatively same, we have presented the results for isotropic case.

\section{Conclusion} \label{sec4}
We propose a graphene/gapped tilted Dirac material/graphene junction which may 
exhibit valley Seebeck effect when the middle region is irradiated with off-resonant circularly 
polarized light. The effect arises due to unequal gaps at the two valleys caused by the combination 
of Semenoff mass $M$ and photoinduced mass $\zeta \Delta$, where $\zeta$ is the valley index. 
Valley polarized thermoelectric properties arise in the device owing to unequal transmission 
probabilities in the conducting channels of the non-degenerate valleys. We have studied the 
valley caloritronics of this junction device in a systematic framework and compared the results 
with the corresponding charge caloritronics. In particular, we have studied the dependence of 
chemical potential $\mu$ and the role of a tunable gap parameter $\eta=M/\Delta$ in the electrical 
and thermal conductances, Seebeck coefficient (thermopower) and figure of merit of 
this junction. Since the the renormalized radiation amplitude  $\beta    
  (=eA_0a/\hbar) \sim 0.1 \ll 1$  
 under 
off-resonant approximation (see Appendix A), the contributions from non-zero order Floquet sidebands 
in the transport properties of the system have been neglected.

The valley polarized electrical conductance $G_v$ attains a maximum and then decreases 
asymptotically to zero while the total charge conductance $G_c$ increases monotonically 
with chemical potential ($\mu$). Furthermore, $G_v$ increases with $\eta$ for $\eta<1$ and 
decreases with $\eta$ for $\eta>1$, while $G_c$ 
shows a decreasing trend with $\eta$. The electrical thermal conductance $\kappa^{\rm el}$ 
as a function of $\eta$ and $\mu$ shows almost similar behavior as charge conductance, as 
it proportional 
to the amount of heat energy carried by the charge carriers.
Both valley ($S_v$) and charge Seebeck coefficients ($S_c$) attain maximum values at 
$\mu\sim \pm 0.04$ eV which is roughly 
independent of $\eta$. But, $S_v$ increases with $\eta$, while $S_c$ shows non-monotonic nature for 
$|\mu|<0.08 $ eV. For $|\mu|<0.08 $ eV, the $S_c$ shows minimum values at $\eta=1$ and starts 
to gain weight as we decrease (for $\eta<1$) or increase $\eta$ (for $\eta>1$). For $|\mu|>0.08 $ eV, 
an increase in $\eta$ leads to a small enhancement in $S_c$ .

Since the ratio of $G$ and $\kappa$ does not show any significant change with $\eta$, 
the figure of merit as a function of $\eta$ shows a variation  similar to square of thermopower 
and its maximum value is obtained at $\mu=\pm0.06$ eV. We have also analyzed the effect of tilting 
in the thermoelectric properties. The tilt parameter  $v_t$ reduces the effective transmission through 
the junction, thereby diminishing all the charge and valley polarized quantities.

The exploitation of valley thermoemf for thermoelectric power generation may serve as a 
new development in the field  of valley caloritronics.  Since the photoinduced mass 
$\Delta$  and Semenoff mass $M$ can be adjusted by varying the intensity of the light source 
and the strength of inversion symmetry-breaking electric field respectively, the tuning of 
the gap parameter $\eta$ may  be achievable in an experimental setup.

\begin{center}
{\bf ACKNOWLEDGEMENTS}
\end{center}
P. Kapri thanks Department of Physics, IIT Kanpur, India for financial support.

\appendix

\section{Floquet Hamiltonian of a tilted Dirac material subjected to circularly 
polarized radiation}
The Hamiltonian for quasiparticles with massive tilted anisotropic Dirac dispersion in the vicinity 
of two independent Dirac points in materials like borophene  or quinoid graphene
is given by
\cite{Zabolotskiy,borophane,Goerbig1,ezawa}
\begin{equation}
H_B({\bf q}) = \zeta\hbar[v_x \sigma_x q_x + \zeta v_y \sigma_y q_y  +  v_t \sigma_{0} q_y ]+
M \sigma_z,
\label{eq4}
\end{equation}
where $\sigma_x$, $\sigma_y$ are the Pauli matrices,
$\sigma_{0}$ is the $2\times2$ identity matrix and 
$\zeta = \pm $ denotes the two independent Dirac points. $M\sigma_z$ is the mass term due to 
different onsite energies $(\pm M)$ of the two sublattices.
The energy dispersion and the corresponding wave
functions associated with the Hamiltonian in Eq. (\ref{eq4}) are given by
\begin{equation}
E_{\lambda, \zeta}({\bf q}) = 
\hbar  \zeta v_t q \sin\theta + \lambda \sqrt{M^2 +[\hbar q \Lambda(\theta)]^2}
\end{equation}
and
\begin{equation}
\Psi_{B}^{\lambda,\zeta}({\bf r}) = \frac{e^{i {\bf q} \cdot {\bf r} }}{\sqrt{2}}
\left(\begin{array}{c}
1   \\
\frac{\zeta \hbar q \Lambda(\theta)
e^{i\zeta \delta } }{M +
\lambda \sqrt{M^2 + [\hbar q \Lambda(\theta)]^2}}
\end{array} \right),
\end{equation}
where $\delta = \tan^{-1}[v_y q_y/(v_x q_x)]$,
$\theta = \tan^{-1}(q_y/q_x)$ and $\Lambda(\theta) =
\sqrt{(v_x \cos\theta)^2 + (v_y \sin\theta)^2 }$ and $\lambda = \pm $ denotes the
conduction and valence bands, respectively.
Note that $v_t \neq 0$ term tilts the Dirac spectrum and is responsible for the
electron-hole symmetry breaking, even for $v_x=v_y$ case.
 
Considering the borophene sheet is illuminated normally by intense circularly polarized 
electromagnetic radiation.
The vector potential corresponding to the circularly polarized radiation is given by
${\bf A}(t) = A_0 (\hat {\bf i} \sin \omega t + \hat {\bf j} \cos \omega t)$, where
$A_0 = E_0/\omega$ with $E_0$ being the amplitude of the electric field vector and
$\omega $ is the frequency of the radiation.  
The vector potential is time-periodic since $ {\bf A}(t + T_\omega) = {\bf A}(t) $,
with the time-period $T_\omega = 2\pi/\omega$. 

The time-periodic Hamiltonian in presence of the electromagnetic radiation 
is given by
\begin{equation}
H_B({\bf q},t) = \zeta\hbar[v_x \sigma_x Q_x(t) + \zeta v_y \sigma_y Q_y(t) 
+  v_t \sigma_{0} Q_y(t)] + M\sigma_z,
\end{equation}
where $ Q_i = q_i + eA_i(t)/\hbar$ with $i=x,y$. 
It is known that a gap in the Dirac spectrum can be induced in graphene, 
on the surface states of topological insulator, silicene, semi-Dirac systems, 
MoS${}_2$ etc by off-resonant radiation. The off-resonant condition is achieved when the 
photon energy ($\hbar \omega$) is much higher than the band width (6$\tau $ with $\tau $ 
being the nearest-neighbor hopping energy) of the undriven borophene.
In the off-resonant condition, the band structure is modified by the second-order
virtual photon absorption-emission processes.
The effective time-independent Floquet Hamiltonian in the off-resonant limit  can
be expressed as \cite{off-resonant,off-resonant1,off-resonant2}
\begin{equation}
H_F({\bf q}) \simeq H_B({\bf q}) + \frac{[H_{-1}({\bf q}), H_{+1}({\bf q})]}{\hbar\omega},
\end{equation}
where the terms in the commutator are the Fourier components of $ H({\bf q},t)$,
\begin{equation}
H_{\pm 1}({\bf q}) = \frac{1}{T_\omega} \int_{0}^{T_\omega} dt e^{ \mp i \omega t} H({\bf q},t).
\label{eq8}
\end{equation}
Using Eq. (\ref{eq8}) we find the commutator $[H_{-1}, H_{+1}]$ as given below,
\begin{equation}
\frac{[H_{-1}({\bf q}), H_{+1}({\bf q})]}{\hbar\omega} = 
\frac {\zeta e^2 A_0^2 v_x v_y}{\hbar \omega} \sigma_z = 
\zeta \Delta \sigma_z,
\end{equation}
where $ \Delta =  (e A_0)^2 v_x v_y/(\hbar \omega)$ is the 
gap at the Dirac points,
an experimentally tunable parameter. 
The gap parameter $\Delta$ does not depend on the tilt parameter $v_t$.

 Here, we would like to mention that the scattering by the Floquet side 
bands are neglected in our study due to the off-resonant condition of light. 
For off-resonant light, the $n$th ($n\neq0$) order Floquet sidebands are separated 
from zeroth order bands (static modes) by large quasienergies ($\sim n\hbar \omega$). 
As discussed in \cite{Kitagawa}, the inelastic scatterings, i.e., photon absorptions and 
emissions between the sidebands are suppressed by a factor of $\beta^2$, 
where $\beta=e A_0 a/\hbar$ is the
renormalized radiation amplitude, with $a$ being the lattice constant. 
Also, the transmission coefficients for sidebands of order $n\neq0$ are small 
($\sim \mathcal{O}(\beta^{2n}))$. 
In our system, the value of $\beta$ for the chosen parameters to evaluate $\Delta$ 
is $\sim 0.1$. Thus, the modification in transmission probability due to 
the scattering by the Floquet side bands is negligibly small.

\section{Transmission probability}
\label{ATP}
In this appendix, we provide the derivation of the transmission probability 
of the electron along with some plots of transmission probability (as a function of incident angle) 
which are required to justify the results presented in Figs. (\ref{FigTEG}) and (\ref{FigTEB}).

The wave functions in the three different regions, $\Psi_{1}(x,y)$, $\Psi_{2}(x,y)$ and 
$\Psi_{3}(x,y)$ will have the same $y$-dependence: $\Psi_{i}(x,y) = \Psi_{i}(x) e^{ik_y y}$
with $i=1,2,3$.
The wave functions, $\Psi_{1}(x)$, $\Psi_{2}(x)$ and $\Psi_{3}(x)$
for the three different regions for $A$ and $B$ sublattices can be written 
in the following forms: for $x<0$,
\begin{eqnarray}
\label{eq7}
\Psi_{1}^\zeta(x) & = & 
\left(\begin{array}{c}
e^{ik_{x}  x}  \\
\zeta e^{i k_x x + i \zeta \phi} \end{array} \right) 
+ r_\zeta
\left(\begin{array}{c}
e^{-ik_x  x}   \\
- \zeta e^{-ik_x x -i \zeta \phi} \end{array} \right),
\end{eqnarray}
for $ 0 <x <L$
\begin{eqnarray}
\label{eq7}
\Psi_{2}^\zeta(x) & = & a_\zeta 
\left(\begin{array}{c}
e^{iq_{x}  x}  \\
\zeta p_{\zeta} e^{i q_x x + i \zeta \delta } \end{array} \right)
+ b_\zeta 
\left(\begin{array}{c}
e^{-iq_x  x}   \\
- \zeta p_{\zeta} e^{-i q_x x - i \zeta \delta } \end{array} \right)
\end{eqnarray}
and for $x>L$
\begin{eqnarray}
\label{eq7}
\Psi_{3}^\zeta(x) & = & t_\zeta 
\left(\begin{array}{c}
e^{ik_x  x}  \\
\zeta e^{i k_x x + i \zeta \phi} \end{array} \right).
\end{eqnarray}
Here the expression for $p_{\zeta}$ is given by
\begin{eqnarray}
\label{eqp}
p_{\zeta} & = & 
\frac{\hbar q \Lambda(\theta)  }
{\Delta_\zeta + \sqrt{\Delta_\zeta^2 + [\hbar q \Lambda(\theta)]^2}}.
\end{eqnarray}
The valley dependent reflection amplitude $r_\zeta$ and the 
transmission amplitude $t_\zeta$ are obtained by matching the wave functions at
the interfaces $x=0$ and $x=L$:
\begin{eqnarray}
\Psi_{1}^\zeta(x=0) = \Psi_{2}^\zeta(x=0); \hspace{0.in} 
\Psi_{2}^\zeta(x=L) = \Psi_{3}^\zeta(x=L).
\end{eqnarray} 
From the above conditions, the valley-dependent transmission probability 
$ \mathcal T_\zeta(\epsilon,\phi) = |t_\zeta(\epsilon,\phi)|^2$ 
is obtained as
\begin{widetext}
\begin{equation}
\label{eqT}
\mathcal T_\zeta(\epsilon,\phi) = \frac{4p_{\zeta}^2 \cos^2\phi \; (1+\cos2 \delta)}
{(2\sqrt{2} p_{\zeta} \cos \phi \cos\delta)^2 + 
[1 + p_{\zeta}^2 - 2p_{\zeta}\cos(\phi-\delta)]
[1 + p_{\zeta}^2 + 2p_{\zeta}\cos(\phi + \delta)][1- \cos(2q L  \cos\theta)] }.
\end{equation}
\end{widetext}
\begin{figure}
\begin{center}
\includegraphics[width=89.5mm, height=45mm]{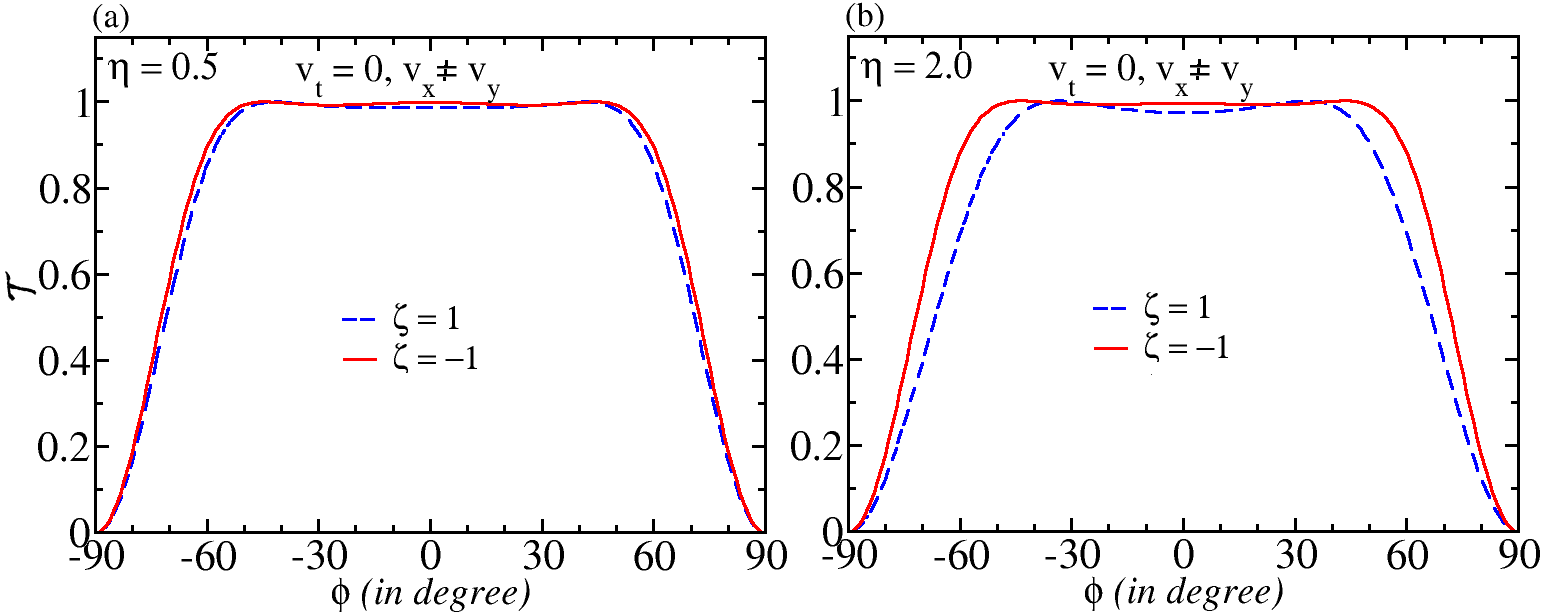}
\caption{ Transmission probability $\mathcal T_\zeta(\epsilon,\phi)$ as a function of
incident angle $\phi$ for a) $\eta = 0.5$, b) $\eta=2$ when $v_t=0$, $v_x\neq v_y$.}
\label{figTNG}
\end{center}
\end{figure}
\begin{figure}
\begin{center}
\includegraphics[width=89.5mm, height=45mm]{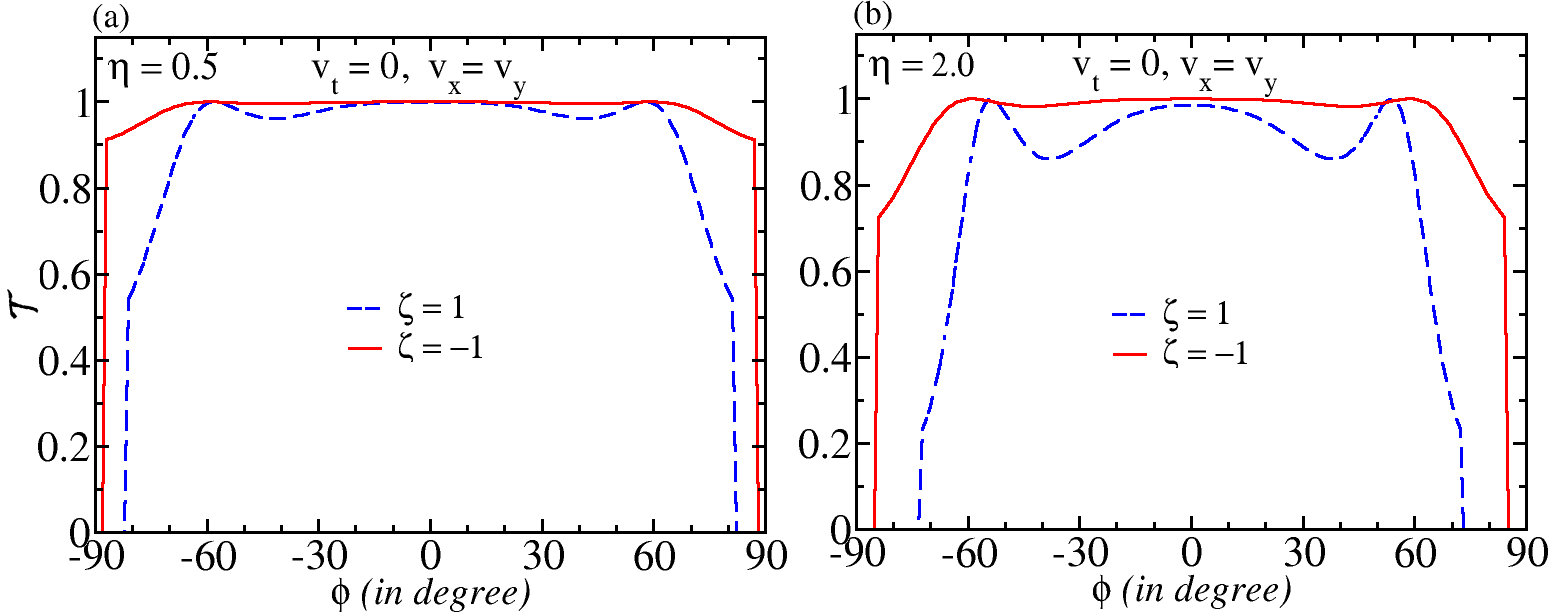}
\caption{Transmission probability $\mathcal T_\zeta(\epsilon,\phi)$ as a function 
of incident angle $\phi$ for a) $\eta = 0.5$, b) $\eta = 2 $ when $v_t=0$,  $v_x= v_y=v_F$.}
\label{figTG}
\end{center}
\end{figure}

Here, it should be noted that there is no mechanism present in the junction that
mixes states of opposite valleys. The system can be reduced to a gapless single
graphene sheet by setting $\Delta_\zeta = 0, v_t = 0$ and $v_x = v_y = v_F$. 
In this limiting case, it can be easily 
checked that $\mathcal T_\zeta(\epsilon, \phi) = 1$.

\begin{figure}
\begin{center}
\includegraphics[width=89.5mm, height=45mm]{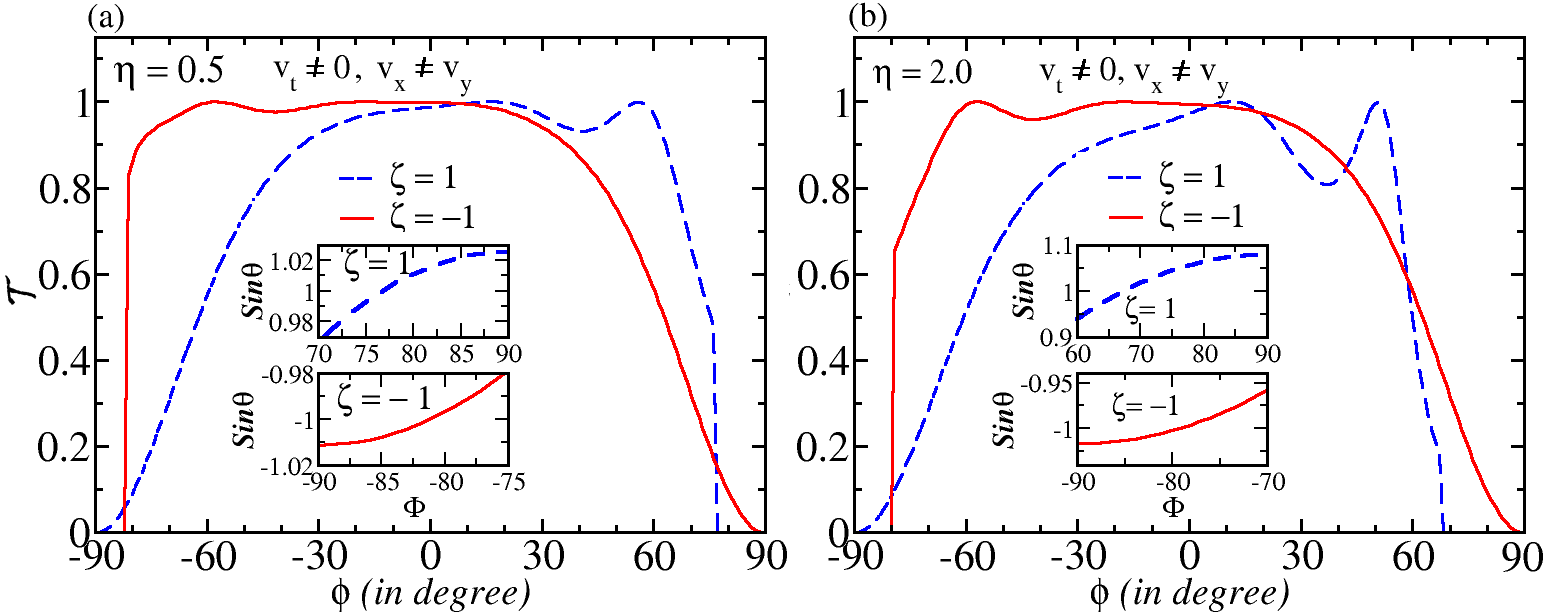}
\caption{Transmission probability $\mathcal T_\zeta(\epsilon,\phi)$ as a function 
of incident angle $\phi$ for a) $\eta = 0.5$, b) $\eta = 2 $ when $v_t \neq 0$, 
$v_x\neq v_y$.}
\label{figTB}
\end{center}
\end{figure}

To understand the behavior of $\mathcal T_\zeta(\epsilon)$ (Figs. (\ref{FigTEG}) and (\ref{FigTEB})), 
plots for $\mathcal T_\zeta(\epsilon,\phi)$ for different conditions as a function
of the incident angle $\phi$ for a fixed energy $\epsilon = 0.5 $ eV and
$L=50 $ nm are shown in Figs. (\ref{figTNG}), (\ref{figTG}) and (\ref{figTB}).

Figures (\ref{figTNG}) and (\ref{figTG}) show plots of $\mathcal T_\zeta(\epsilon,\phi)$ as a 
function of $\phi$ for $(i)$ $v_t=0, v_x \neq v_y$ and  $(ii)$  $v_t=0, v_x=v_y=v_F$ for two values of 
$\eta$ with $\Delta$ fixed at $0.05$ eV.
On the other hand, Fig. (\ref{figTB}) shows plot of $\mathcal T_\zeta(\epsilon,\phi)$ 
as a function of $\phi$ for $v_t \neq 0, v_x \neq v_y$.
All the figures show that the transmission probability
is close to unity around the normal incidence ($\phi\to 0$) for both the valleys.
This is manifestation of perfect tunneling
when incident wave vector is normal to the interface.
Figure (\ref{figTNG}) shows that the transmission is allowed over the full range of 
incident angle ($-\pi/2 \leq \phi \leq \pi/2$), whereas  for Fig. (\ref{figTG}), the 
transmission is restricted below the lower critical angle and above the upper critical 
angle. Further, in Fig. (\ref{figTB}), the transmission probability 
for $\zeta = +/-$ ceases to zero above/below some critical incident angles.
This is because of $\sin\theta > 1$, as shown in the
inset of the Fig. (\ref{figTB}), for which above or below the critical angles the wave 
vector in the middle region becomes complex which leads to evanescent wave.  
Figures (\ref{figTG}) and (\ref{figTB}) reveal that the
allowed range of incident angle for $\zeta = - 1$ is bigger than that of the $\zeta=1$, as 
for the latter one, the band gap is wider.
It it is clear that the value of critical angles depend on all the three velocities
$v_x$, $v_y$ and $v_t$.  
Similar critical angles exist in other semiconductor junction devices
\cite{angle,angle1}.

\section{Theory of Thermoelectricity} \label{ATTE}
In this appendix, we present the derivation of thermopower and electron's thermal conductance
in terms of the transmission probability.

Assuming that the graphene leads are independent electron reservoirs,
the chemical potential and the temperature of the left/right graphene leads 
are $\mu_{L/R}$ and $T_{L/R}$, respectively. 
The population of electrons at the left/right leads is described by the 
Fermi-Dirac distribution function
$f_{L/R} = f(\mu_{L/R},T_{L/R}) = [1+e^{(\epsilon - \mu_{L/R})/(k_BT_{L/R})}]^{-1}$.  
Employing the Landauer-Buttiker formalism in quasi-ballistic regime, the valley-dependent 
charge current is given by
\begin{equation}
\label{elec-current}
I_{\zeta} = \frac{2e}{h} \int \limits_{-\pi/2}^{\pi/2} d\phi \cos\phi
\int \limits_{-\infty}^{\infty} N(\epsilon) \mathcal T_{\zeta}(\epsilon,\phi)(f_{L} - f_{R}) d\epsilon,
\end{equation}
where $N(\epsilon) = W|\epsilon|/(\pi\hbar v_F)$ is the energy dependent
number of transverse modes in the graphene sheet of width $W$ \cite{Beenakker}. 
Here it has been used that $\mathcal T_{L,{\zeta}}(\epsilon,\phi) 
= \mathcal T_{R,{\zeta}}(\epsilon,\phi) 
= \mathcal T_{\zeta}(\epsilon,\phi)$ 
with $T_{L,\zeta}(\epsilon,\phi)$ $(T_{R,\zeta}(\epsilon,\phi))$ is the transmission probability of an electron
with energy $\epsilon$ and incidence angle $\phi$ from left (right) graphene leads.

In absence of any external bias voltage $(V_B)$, the chemical potentials 
of the two leads are taken to be the same as $\mu_L = \mu_R = \mu$.
Due to the applied temperature difference $(dT)$ between the two leads, 
there will be a small voltage difference $(dV)$ between the leads. 
The currents induced by $dT$ and $dV$ are given by
$(dI_{\zeta})_T = I_{\zeta}(\mu, T; \mu, T+dT)$ and 
$(dI_{\zeta})_V = I_{\zeta}(\mu, T; \mu + edV, T)$,  
where the currents $I_{\zeta}(\mu, T; \mu, T+dT)$ and 
$I_{\zeta}(\mu, T; \mu + edV, T)$ can be calculated 
from Eq. (\ref{elec-current}).
Since in an open circuit condition, the current cannot flow, one can write
\begin{eqnarray}
\label{elec-open}
dI_{\zeta} = (dI_{\zeta})_T + (dI_{\zeta})_V = 0.
\end{eqnarray}
Expanding the Fermi-Dirac distribution functions in  
Eq. (\ref{elec-current}) and Eq. (\ref{elec-open}) in the linear response regime, i.e.,  
up to the first order terms in $dV$ and $dT$, 
one can get the valley resolved Seebeck coefficient $S_{\zeta}$ as
\begin{equation}
\label{EqnS}
S_{\zeta} = -\frac{dV}{dT}\Big|_{dI_{\zeta}=0} = - \frac{L^{(1)}_{\zeta}}{eTL^{(0)}_{\zeta}},
\end{equation}
where the kinetic coefficients $L^{(\alpha)}_{\zeta}$ for the quasi-ballistic transport regime are given by
\begin{equation}\label{lalpha}
L^{(\alpha)}_{\zeta} =\int_{-\pi/2}^{\pi/2}d\phi \cos\phi 
\int_{-\infty}^{\infty} \mathcal T_{\zeta}(\epsilon,\phi )N(\epsilon)(\epsilon - \mu)^{\alpha}
(-\frac{\partial f}{\partial \epsilon})d\epsilon 
\end{equation}
with $\alpha = 0,1,2$.

The flow of electrons can also transport thermal energy through the junction, 
which is responsible for the thermal current.
The electron's thermal current is the energy current carried by
electrons traveling between leads driven by $dT=T_R-T_L$ and $dV = (\mu_R-\mu_L)/e$.
Analogous to the charge current, the electron's 
valley resolved thermal current can be written as
\begin{eqnarray}
\label{therm-current}
J^{\rm el}_{\zeta} = \frac{2}{h} \int \limits_{-\pi/2}^{\pi/2}d\phi \cos\phi
\int \limits_{-\infty}^{\infty}N(\epsilon) \mathcal T_{\zeta} (\epsilon,\phi)
(\epsilon - \mu)(f_L - f_R) d\epsilon. \nonumber \\
\end{eqnarray}

Analogous to the charge current driven by $dT$ and $dV$, the electron's
valley resolved thermal current can be written as
\begin{equation}
dJ^{\rm el}_{\zeta} = (dJ^{\rm el}_{\zeta})_T + (dJ^{\rm el}_{\zeta})_{V},
\end{equation}
where $(dJ^{\rm el}_{\zeta})_T = J^{\rm el}_{\zeta}(\mu,T; \mu,T+dT)$ and
$(dJ^{\rm el}_{\zeta})_V = J^{\rm el}_{\zeta}(\mu, T; \mu + edV,T)$.
Note that $dV$ is generated by the Seebeck
effect due to the temperature difference $dT$. Both
$J^{\rm el}_{\zeta}(\mu, T; \mu, T + dT)$ and $J^{\rm el}_{\zeta}(\mu, T; \mu + edV, T)$ can be
calculated using Eq. (\ref{therm-current}).
Similarly, the electron's thermal conductance, $k^{\rm el}_{\zeta} = dJ^{\rm el}_{\zeta}/dT$ has
two components:
\begin{equation}
k^{\rm el}_{\zeta} = (k^{\rm el}_{\zeta})_T + (k^{\rm el}_{\zeta})_V,
\end{equation}
where $(k^{\rm el}_{\zeta})_T = (dJ^{el}_{\zeta})_{T}/dT$ and 
$(k^{\rm el}_{\zeta})_V = (dJ^{\rm el}_{\zeta})_V/dT$
are the portions of the electron's thermal conductance driven by $dT$
and $dV$ respectively. The electron's valley resolved thermal conductance can be expressed 
in terms of the kinetic coefficients $L^{(\alpha)}_{\zeta}$ (as given in \ref{lalpha})  \cite{Mawrie,Liu} as
\begin{equation}
\label{eq48}
k^{\rm el}_{\zeta} = \frac{2}{h}\frac{L^{(2)}_{\zeta}}{T} +  \frac{2e}{h}L^{(1)}_{\zeta}S_{\zeta}.
\end{equation}

\end{document}